\documentclass[aps,pra,a4paper,floatfix,twocolumn,longbibliography,groupedaddress,superscriptaddress]{revtex4-1}

\usepackage{amsmath}
\usepackage{enumitem}
\usepackage{amssymb}
\usepackage{amsfonts}
\usepackage{braket}
\usepackage{dsfont}
\usepackage{MnSymbol} 

\usepackage{color}
\newcommand{\revision}[1]{\textcolor{black} { #1 }}
\usepackage[dvipsnames]{xcolor}
\usepackage{amsmath}
\usepackage{tikz}

\usepackage{graphicx, import}
\usepackage{soul}
\usepackage[utf8]{inputenc}
\usepackage{caption}
\usepackage{subcaption}
\usepackage{ragged2e}
\usepackage{hyperref}
\usepackage{parskip}

\begin{document}
\title{Let Quantum Neural Networks Choose Their Own Frequencies}
\author{Ben Jaderberg}
\author{Antonio A. Gentile}
\affiliation{PASQAL, 7 rue Léonard de Vinci, 91300 Massy, France}
\author{Youssef Achari Berrada}
\author{Elvira Shishenina}
\affiliation{BMW Group, 80788 Munich, Germany}
\author{Vincent E. Elfving}
\affiliation{PASQAL, 7 rue Léonard de Vinci, 91300 Massy, France}
\date{\today}

\begin{abstract}
Parameterized quantum circuits as machine learning models are typically well described by their representation as a partial Fourier series of the input features, with frequencies uniquely determined by the feature map's generator Hamiltonians. Ordinarily, these data-encoding generators are chosen in advance, fixing the space of functions that can be represented. In this work we consider a generalization of quantum models to include a set of trainable parameters in the generator, leading to a trainable-frequency (TF) quantum model. We numerically demonstrate how TF models can learn generators with desirable properties for solving the task at hand, including non-regularly spaced frequencies in their spectra and flexible spectral richness. Finally, we showcase the real-world effectiveness of our approach, demonstrating an improved accuracy in solving the Navier-Stokes equations using a TF model with only a single parameter added to each encoding operation. Since TF models encompass conventional fixed-frequency models, they may offer a sensible default choice for variational quantum machine learning.

\end{abstract}

\maketitle

\section{Introduction} \label{sec:introduction}

The field of quantum machine learning (QML) remains a promising application for quantum computers. In the fault-tolerant era, the prospect of quantum advantage is spearheaded by the exponential speedups in solving linear systems of equations \cite{harrow2009quantum}, learning distributions \cite{pirnay2022super, liu2021rigorous} and topological data analysis~\cite{lloyd2016quantum}. Yet the arrival of large fault-tolerant quantum computers is not anticipated in the next decade, reducing the practical impact of such algorithms today.

\begin{figure}[h!]
\includegraphics[width=\columnwidth]{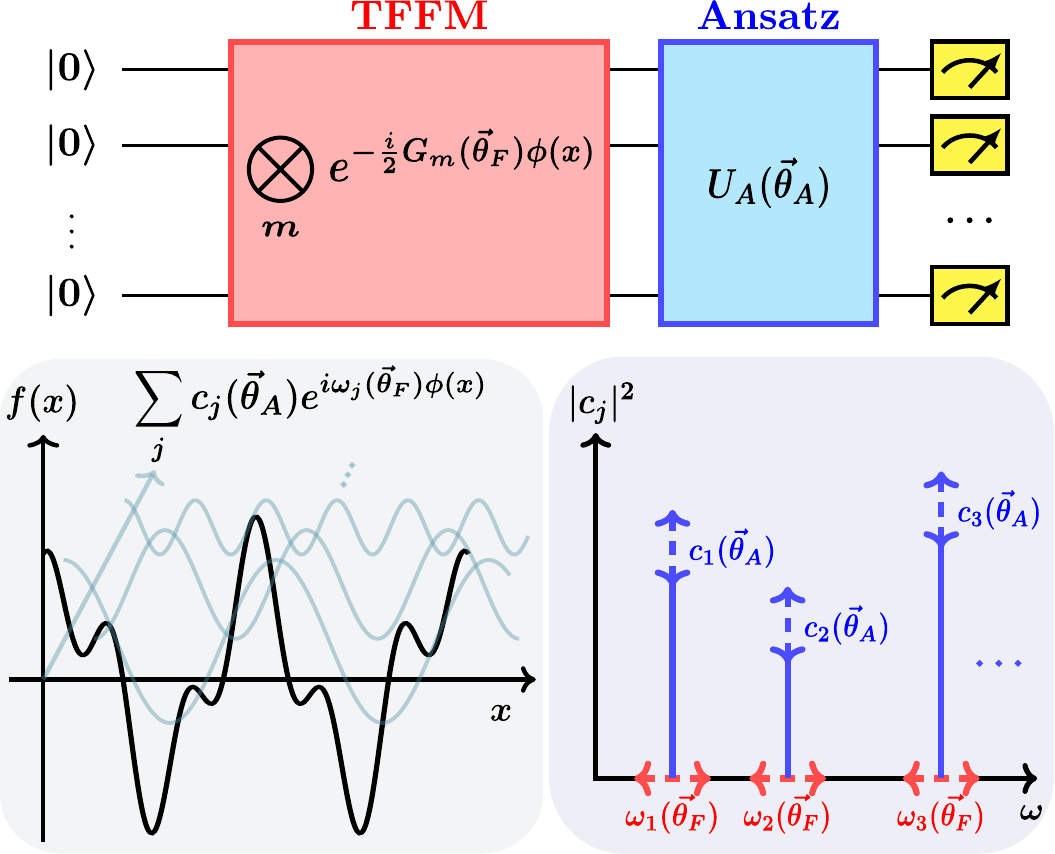}
    \captionsetup{justification=raggedright} 
    \caption{An overview of the concepts discussed in this paper. Top: we introduce a parameterized quantum circuit in which the generator of the data-encoding block is a function of trainable parameters $\vec{\theta}_F$, alongside the standard trainable variational ansatz. Bottom left: the output of such a model is a Fourier-like sum over different individual modes. Bottom right: in conventional quantum models, tuning the ansatz parameters $\vec{\theta}_A$ allows the coefficients of each mode to be changed. By using a trainable-frequency feature map (TFFM), tuning $\vec{\theta}_F$ leads to a quantum model in which the frequencies of each mode can also be trained. }
    \label{fig:overview}
\end{figure}

One approach to solving relevant problems in machine learning with today's quantum computers is through the use of parameterized quantum circuits (PQCs)~\cite{benedetti2019parameterized, mitarai2018quantum}, which have been applied to a variety of use cases~\cite{farhi2018classification,havlicek2019supervised,otterbach2017unsupervised,bausch2020recurrent,guan2021quantum, heim2021quantum,kyriienko2022protocols}. A PQC consists of quantum feature maps (FMs) $\hat{U}_F(\vec{x})$, which encode an input $\vec{x}$ into the Hilbert space, and variational ans{\"a}tze $\hat{U}_A(\vec{\theta}_A)$ which contain trainable parameters. Previously, it was shown that the measured output of many variational QML models can be mathematically represented as a partial Fourier series in the network inputs \cite{schuld2021effect}, leading to a range of new insights~\cite{heimann2022learning, peters2022generalization, kyriienko2021generalized}. Most strikingly, it follows that the set of frequencies $\Omega$ appearing in the Fourier series  are uniquely determined by the eigenvalues of the generator Hamiltonian of the quantum FM, while the series coefficients are tuned by the variational parameters $\vec{\theta}_A$. 

Conventionally, a specific generator is chosen beforehand, such that the model frequencies are fixed throughout training. In theory this is not a problem since, by choosing a generator that produces regularly spaced frequencies, the basis functions of the Fourier series form an orthogonal basis set. This ensures that asymptotically large fixed-frequency (FF) quantum models are universal function approximators~\cite{schuld2021effect}. Yet in reality, finite-sized quantum computers will permit models with only a finite number of frequencies. Thus, in practice, great importance should be placed on the \emph{choice} of basis functions, for which the orthogonal convention may not be the best.

This raises an additional complexity: what is the optimal choice of basis functions? Indeed for many problems, it is not obvious what this would be without prior knowledge of the solution. Here we address this issue by exploring a natural extension of quantum models in which an additional set of trainable parameters $\vec{\theta}_{F}$ is included in the FM generator. This simple idea has a significant impact on the effectiveness of quantum models, allowing the generator eigenspectrum to change over the course of training in a direction that minimises the objective loss. This in turn creates a quantum model with trainable frequencies as visualised in Figure \ref{fig:overview}. In a quantum circuit learning~\cite{mitarai2018quantum} setting, we numerically demonstrate cases in which trainable-frequency (TF) models can learn a set of basis functions that better solves the task at hand, such as when the solution has a spectral decomposition with non-regularly spaced frequencies.

Furthermore, we show that an improvement is realisable on more advanced learning tasks. We train quantum models with the differentiable quantum circuits (DQC) algorithm~\cite{kyriienko2021solving} to learn the solution to the two-dimensional (2D) time-dependent Navier-Stokes differential equations, \revision{a family of equations that has proven challenging to solve with quantum models previously~\cite{sedykh2023quantum}}. For the problem of predicting the wake flow of fluid passing a circular cylinder, a TF quantum model achieves lower loss and better predictive accuracy than the equivalent FF model. Overall, our results raise the prospect that TF quantum models could improve performance for other near-term QML problems.

\section{Previous works}\label{sec:previous_works}

The idea of including trainable parameters in the feature map generator is present in some previous works. In a study of FM input redundancy, Gil Vidal et al.~\cite{gil2020input} hypothesise that a variational input encoding strategy may improve the expressiveness of quantum models, followed by limited experiments~\cite{lei2020comparisons}. Other works also suggest that encoding unitaries with trainable weights can reduce circuit depths of quantum models, as discussed for single-qubit classifiers~\cite{perez2020data} and quantum convolutional neural networks~\cite{ovalle2023quantum}. Nevertheless, our work is different due to contributions demonstrating (a) an analysis of the effect of trainable generators on the Fourier modes of quantum models, (b) evidence of specific spectral features in data for which TF models offer an advantage over FF models and (c) direct comparison between FF and TF models for a practically relevant learning problem. 

In the language of quantum kernels \cite{havlicek2019supervised, henry2021quantum, paine2023quantum}, several recent works use the term ``trainable feature map'' to describe the application of unitaries with trainable parameters on data already encoded into the Hilbert space~\cite{john2023optimizing, gentinetta2023quantum}. Such a distinction is necessary because quantum kernel models often contain no trainable parameters at all. However, the trainable parameters of these feature maps do not apply directly to the generator Hamiltonian. As discussed in section \ref{sec:method}, this is intrinsically different from our scheme as it does not lead to a model with trainable frequencies. To not confuse the two schemes, here we adopt the wording ``trainable-\emph{frequency} feature map'' and ``trainable-\emph{frequency} models''.

\section{Method}\label{sec:method}

Practically, quantum computing entails the application of sequential operations (e.g., laser pulses) to a physical system of qubits. Yet to understand how these systems can be theoretically manipulated, it is often useful to work at the higher-level framework of linear algebra, from which insights can be translated back to real hardware. This allows studying strategies encompassing digital, analog, and digital-analog paradigms \cite{parrarodriguez2020digitalanalog}. In a more abstract formulation, a broad class of FMs can be described mathematically as the tensor product of an arbitrary number of sub-feature-maps, each represented by the time evolution of a generator Hamiltonian applied to an arbitrary subset of the qubits

\begin{equation}\label{eqn:fm}
    \hat{U}_F(\vec{x}) = \bigotimes_{m} e^{-\frac{i}{2}\hat{G}_m(\gamma_m)\phi(\vec{x})}, 
\end{equation}

where for the sub-feature-map $m$, $\hat{G}_m(\gamma_m)$ is the generator Hamiltonian that depends on non-trainable parameters $\gamma_m$. Furthermore, $\phi(\vec{x}): \mathbb{R}^n \rightarrow \mathbb{R}^n$ is an encoding function that depends on the input features $\vec{x}$. Practically speaking, $\gamma_m$ is typically related to the index of the tensor product space the sub-feature-map is applied to and can be used to set the number of unique frequencies the model has access to. Furthermore, in some cases  $\hat{U}_F(\vec{x})$ can be applied several times across the quantum circuit, interleaved with variational ansatz $\hat{U}_A(\vec{\theta}_A)$ layers, for example in data re-uploading \cite{perez2020data, fan2022compact} and serial feature maps \cite{schuld2021effect}. 

The measured output of a quantum model with a feature map defined in Eq. (\ref{eqn:fm}) can be expressed as a Fourier-type sum in the input dimensions

\begin{equation}\label{eqn:FFFM_model_output}
    f(\vec{x}, \vec{\theta}_A) = \sum_{\vec{\omega}_j \in \Omega} \vec{c}_j(\vec{\theta}_A) e^{i\vec{\omega}_j \cdot \phi(\vec{x})},
\end{equation}

where $\vec{c}_j$ are the coefficients of the multi-dimensional Fourier mode with frequencies $\vec{\omega}_j$. Crucially, the frequency spectrum $\Omega$ of a model is uniquely determined by the eigenvalues of $\hat{G}_m$~\cite{schuld2021effect}. More specifically, let us define the final state produced by the PQC as $\ket{\psi_f}$. If the model is a quantum kernel, in which the output derives from the distance to a reference quantum state $\ket{\psi}$ (e.g., $f(\vec{x}, \vec{\theta}_A) = |\langle \psi | \psi_f \rangle|^2$), then $\Omega$ is explicitly the set of eigenvalues of the composite generator $\hat{G}$ such that $ \hat{U}_F(\vec{x}) = e^{-\frac{i}{2}\hat{G}\phi(\vec{x})}$. When the generators $\hat{G}_m$ commute, the composite generator is simply $\hat{G}=\sum_m G_m$. If the model is a quantum neural network (QNN), where the output is derived from the expectation value of a cost operator $\hat{C}$ (e.g., $f(\vec{x}, \vec{\theta}_A) = \langle \psi_f|\hat{C}| \psi_f \rangle$), then $\Omega$ contains the gaps in the eigenspectrum of $\hat{G}$.

The key insight here is that in such quantum models, a specific feature map generator is chosen in advance. This fixes the frequency spectrum over the course of training, setting predetermined basis functions $e^{i\vec{\omega}_j\cdot\phi(\vec{x})}$ from which the model can construct a solution. For these FF models, only the coefficients $\vec{c}_j$ can be tuned by the variational ansatz during training.

In this work, we replace the FM generator with one that includes trainable parameters $\hat{G}_m(\gamma_m, \vec{\theta}_{F})$. In doing so, the generator eigenspectrum, and thus the model frequencies, can also be tuned over the course of training. For this reason we refer to such feature maps as trainable-frequency feature maps (TFFMs), which in turn create TF quantum models. The output of a TF quantum model will be a Fourier-type sum in which the frequencies of each mode depend explicitly on the parameterization of the feature map. For example, for a QNN the output of the model can be written as

\begin{equation}\label{eqn:TFFM_model_output}
    f(\vec{x}, \vec{\theta}_A, \vec{\theta}_{F}) = \sum_{\vec{\omega}_j \in \Omega} \vec{c}_j(\vec{\theta}_A, \hat{C}) e^{i\vec{\omega}_j(\vec{\theta}_{F})\cdot\phi(\vec{x})},
\end{equation}

Moreover, as $\vec{\theta}_{F}$ is optimized with respect to minimising a loss function $\mathcal{L}$, the introduction of trainable frequencies $\vec{\omega}_j(\vec{\theta}_{F})$ ideally allows the selection of spectral modes that better fit the specific learning task. For gradient-based optimizers, the derivative of parameters in the generator $\frac{\partial \mathcal{L}}{\partial \vec{\theta}_F}$ can be calculated as laid out in Appendix \ref{app:tffm_derivatives}, which for the generators used in our experiments simplifies to the parameter-shift rule (PSR)~\cite{crooks2019gradients, kyriienko2021generalized}.

We note that the idea here is fundamentally different from simply viewing a combination of feature maps and variational ans{\"a}tze (e.g., serial feature maps) as a higher level abstraction containing one unitary $\hat{U}_{\tilde{F}}(\vec{x}, \vec{\theta}) = \hat{U}_F(\vec{x})\hat{U}_A(\vec{\theta})$. The key concept in TF models is that a parameterization is introduced that acts directly on the generator Hamiltonian in the exponent of Eq. (\ref{eqn:fm}). This is what allows a trainable eigenvalue distribution, leading to quantum models with trainable frequencies.

Furthermore, in this work the FF models we consider are those in which the model frequencies are regularly spaced (i.e., integers or integer valued multiples of a base frequency), such that the basis functions form an orthogonal set. This has become the conventional choice in the literature \cite{kubler2021inductive, heim2021quantum, ghosh2022harmonic, skolik2022quantum, hubregtsen2022training, jerbi2023quantum, meyer2023exploiting}, owing to the theoretical grounding that such models are universal function approximators in the asymptotic limit~\cite{schuld2021effect}. However, it should be made clear that a FF model could mimic any TF model if the non-trainable unitaries in the FM were constructed with values corresponding to the final trained values of $\vec{\theta}_F$. The crux, however, is that having knowledge of such values without going through the training process is highly unlikely and might occur only where considerable a-priori knowledge of the solution is available. 

\section{Proof of principle results}\label{sec:pop_results}

The potential advantage of TF quantum models stem their ability to be trained such that their frequency spectra contain non-uniform gaps, producing non-orthogonal basis functions. We demonstrate this effect by first considering a fixed-frequency feature map (FFFM) in which, for simplicity, we restrict $\hat{G}_m$ to single-qubit operators. Overall, we choose a generator Hamiltonian $\hat{G} = \sum_{m=1}^N \gamma_m\hat{Y}^m/2$, where $N$ is the number of qubits, $\hat{Y}^m$ is the Pauli matrix applied to the tensor product space of qubit $m$, $\gamma_m = 1$, and $\phi(\vec{x}) = x$ is a one-dimensional Fourier encoding function. This is the commonly used angle encoding FM~\cite{larose2020robust}, which we use to train a QNN to fit data produced by a cosine series

\begin{equation}\label{eqn:cosine_series}
    y(x) = \frac{1}{|\Omega_d|}\sum_{\omega_d \in \Omega_D} \cos(\omega_dx).
\end{equation}

Here the data function contains a set of frequencies $\Omega_d$, from which $n_d$ data points are generated equally spaced in the domain $\mathcal{D} = [-4\pi, 4\pi]$. The value of $n_d$ is determined by the Nyqvist sampling rate such that $n_d = \lceil 2 |\mathcal{D}| \max(\Omega_d)\rceil$.

\begin{figure*}
  \centering
  \begin{subfigure}[b]{0.32\textwidth}
    \includegraphics[width=\textwidth]{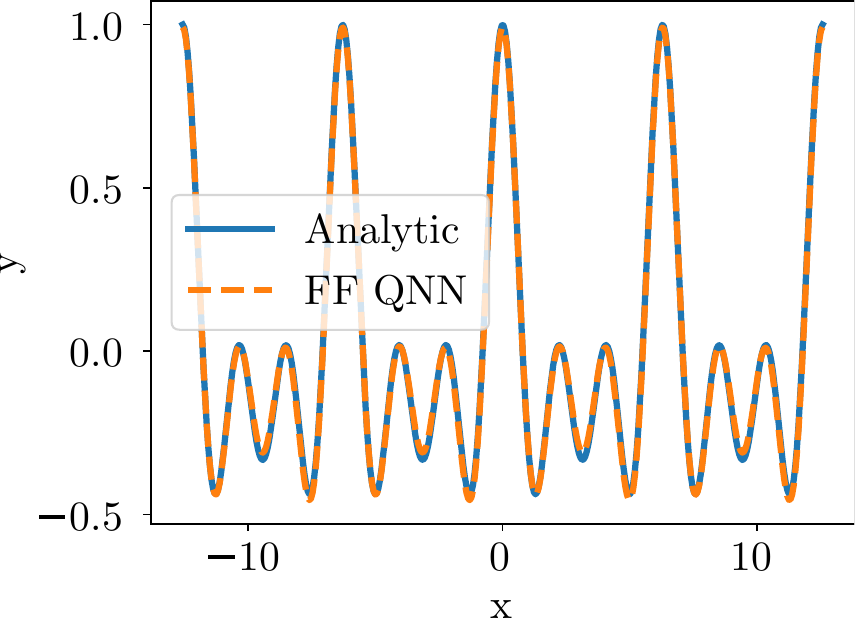}
    \caption{}
    \label{fig:spectral_spacing_a}
  \end{subfigure}
  \hfill
  \begin{subfigure}[b]{0.32\textwidth}
    \includegraphics[width=\textwidth]{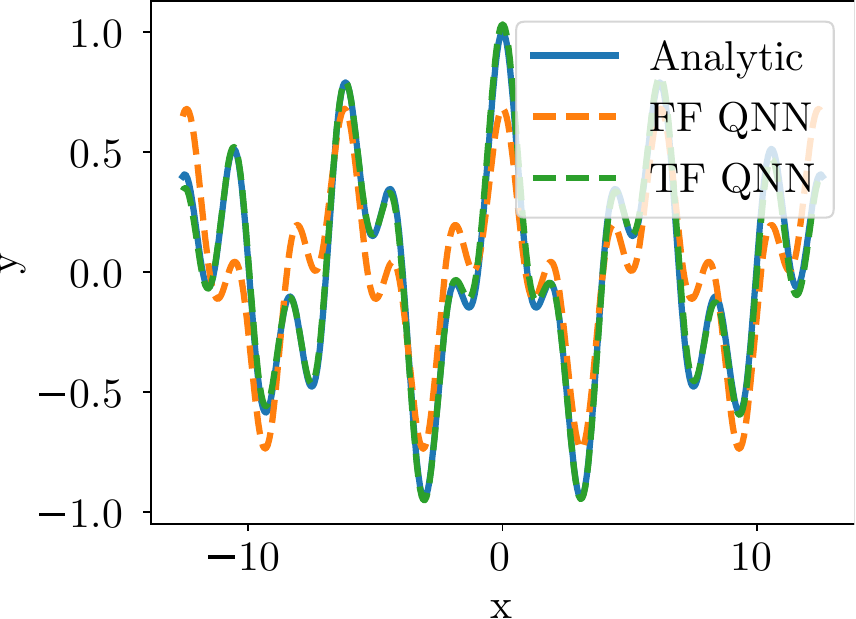}
    \caption{}
    \label{fig:spectral_spacing_b}
  \end{subfigure}
  \hfill
  \begin{subfigure}[b]{0.32\textwidth}
    \includegraphics[width=\textwidth]{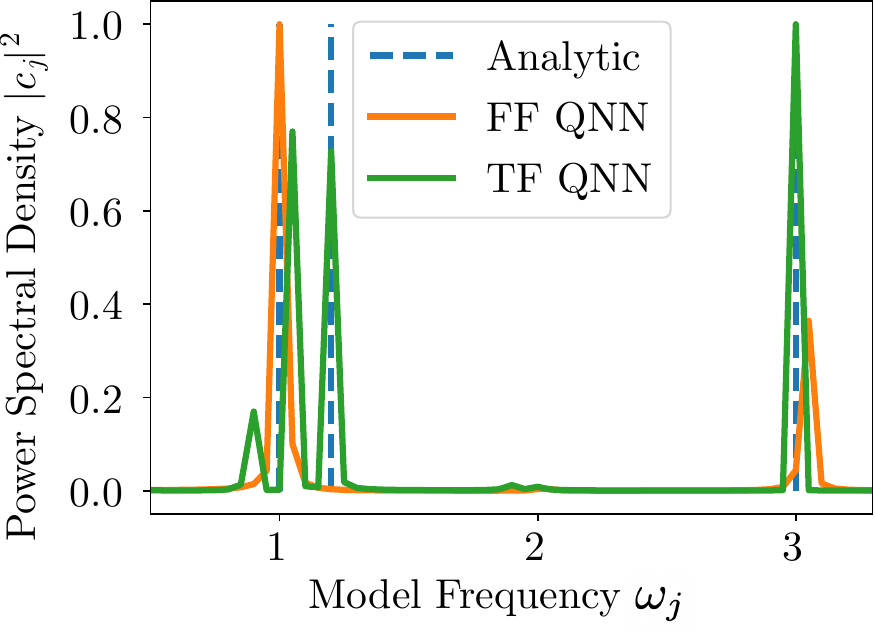}
    \caption{}
    \label{fig:spectral_spacing_c}
  \end{subfigure}
  \caption{Fitting cosine series of different frequencies using fixed-frequency (FF) and trainable-frequency (TF) QNNs. (a) Prediction after training on data with $\Omega_d=\{1, 2, 3\}$ (b) Prediction after training on data with $\Omega_d=\{1, 1.2, 3\}$. (c) Spectra of trained models in (b) as obtained using a discrete Fourier transform. The blue dashed lines indicate frequencies of the data.}
  \label{fig:spectral_spacing_results}
\end{figure*}

Figure \ref{fig:spectral_spacing_results} demonstrates where FF models succeed and fail. In these experiments, a small QNN with $N=3$ qubits and $L=4$ variational ansatz layers is used. More details of the quantum models and training hyperparameters for all experiments can be found in Appendix \ref{app:hyperparameters}. 

The FF QNN defined above is first trained on data with frequencies $\Omega_d=\{1, 2, 3\}$. After training, the prediction of the model is recorded as shown in Figure \ref{fig:spectral_spacing_a}. Here we see that the underlying function can be perfectly learned. To understand why, we note that the set of degenerate eigenvalues of $\hat{G}$ are $\lambda = \{-\frac32, -\frac12, \frac12, \frac32\}$ and thus the unique gaps are $\Delta = \{1, 2, 3\}$. In this case, the natural frequencies of the model $\Delta$ are equal to the frequencies of the data $\Omega_d$, making learning trivial. Furthermore, we find that similar excellent fits are possible for data containing frequencies that differ from the natural model frequencies by a constant factor (e.g., $\Omega_d=\{1.5, 3, 4.5\}$), provided the quantum model is given the trivial classical resource of parameters that can globally scale the input $x$. 

By contrast, Figure \ref{fig:spectral_spacing_b} illustrates how a FF QNN fails to fit data with frequencies $\Omega_d=\{1, 1.2, 3\}$. This occurs because no global scaling of the data can enable the fixed generator eigenspectrum to contain gaps with unequal spacing. In such a setting, no additional training would lead to accurate fitting of the data. Furthermore, no practical number of extra ansatz layers would enable the FF model to fit the data (see section \ref{sec:conclusion} for further discussion), for which we verify up to $L=128$. 

Conversely, a QNN containing a TFFM with the simple parameterization $\hat{G}_{\theta}= \sum_{m=1}^N \theta_m\hat{Y}^m/2$ does significantly better in fitting the data. This is precisely possible because training of the generator parameters converges on the values $\vec{\theta}_{F}=\{0.89, 1.05, 1.04\}$, leading to an eigenvalue spectrum which contains gaps $\Delta_\theta = \{..., 0.95, 1.050, 1.200, 3.000, ...\}$ as shown in Figure \ref{fig:spectral_spacing_c}. In this experiment and all others using TF models, the TFFM parameters are initialised as the unit vector $\vec{\theta}_{F} = \mathds{1}$.

A further advantage of TF models is their flexible spectral richness. For FF models using the previously defined $\hat{G}$, one can pick values $\gamma_m = 1$, $\gamma_m = m$ or $\gamma_m = 2^{(m-1)}$ to produce generators where the number of unique spectral gaps $|\Omega|$ scales respectively as $\mathcal{O}(N), \mathcal{O}(N^2)$ and $\mathcal{O}(2^N)$ respectively. Typically, a practitioner may need to try all of these so-called simple, tower~\cite{kyriienko2021solving} and exponential~\cite{kyriienko2022protocols} FMs, yet a TFFM can be trained to effectively represent any of these. To test this, we again sample from Eq. (\ref{eqn:cosine_series}) to construct seven data sets that contain between one and seven frequencies equally spaced in the range $\omega_d \in [1, 3]$. 

\begin{figure}
    \centering
    \includegraphics[width=0.8\columnwidth]{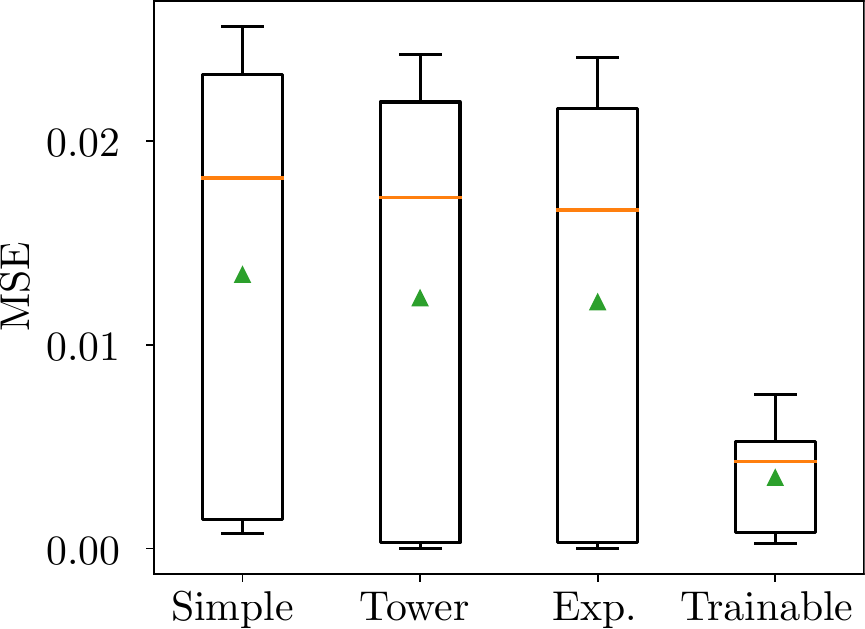}
    \caption{Prediction MSE of simple, tower and exponential FFFMs compared to a TFFM when training on multiple data sets with equally spaced frequencies $\vec{\omega_d} \in [1, 3]$. For each box, the triangle and orange line denote the mean and median respectively.}
    \label{fig:spectral_richness_results}
\end{figure}

Figure \ref{fig:spectral_richness_results} shows the mean squared error (MSE) achieved by each QNN across these data sets. In the best case, the TF and FF  models perform equally well, since each fixed generator produces a specific number of frequencies for which it is well suited for. However, we find that a TF model outperforms FF models in the average and worst-case scenarios. Despite the data having orthogonal basis functions, the FF models have either too few frequencies (e.g., simple FM for data with $|\Omega_d|>3$) or too many (e.g., exponential FM for data with $|\Omega_d|<7$) to perform well across all data sets. Thus, we find that even when the optimal basis functions are orthogonal, TF models can be useful when there is no knowledge of the ideal number of spectral modes of the solution.

\section{Application to fluid dynamics}\label{sec:ns_results}

In this section we demonstrate the impact of TF models on solving problems of practical interest. Specifically, we focus on the DQC algorithm \cite{kyriienko2021solving}, in which a quantum model is trained to find a solution to a partial differential equation (PDE). The PDE to be solved is the incompressible 2D time-dependent Navier-Stokes equations (NSEs), defined as

\begin{align}
\revision{\zeta_x}&\revision{(x,y,t) =} \notag\\
&\frac{\partial u}{\partial t} + u\frac{\partial u}{\partial x} + v\frac{\partial u}{\partial y} - \frac{1}{Re}\left(\frac{\partial^2 u}{\partial x^2} + \frac{\partial^2 u}{\partial y^2}\right) + \frac{\partial p}{\partial x} = 0, \label{eqn:ns_x}\\
\revision{\zeta_y}&\revision{(x,y,t) =} \notag\\
&\frac{\partial v}{\partial t} + u\frac{\partial v}{\partial x} + v\frac{\partial v}{\partial y} - \frac{1}{Re}\left(\frac{\partial^2 v}{\partial x^2} + \frac{\partial^2 v}{\partial y^2}\right) + \frac{\partial p}{\partial y} = 0, \label{eqn:ns_y}
\end{align}

where $u$ and $v$ are the velocity components in the $x$ and $y$ directions respectively, $p$ is the pressure and $Re$ is the Reynolds number, which represents the ratio of inertial forces to viscous forces in the fluid.

The goal of this experiment is to train a quantum model to solve the downstream wake flow of fluid moving past a circular cylinder. This is one of the canonical systems of study in physics-informed neural networks, the classical analogue of DQC, due to the vortex shedding patterns and other complex dynamics exhibited even in the laminar regime~\cite{raissi2017physics, rao2020physics, xu2023practical}. A high-resolution data set for $Re=100$ is obtained from \cite{raissi2017physics} for the region $x=1$ to $x=8$ downstream from a cylinder at $x=0$, solved using the NekTar high-order polynomial finite element method (FEM) \cite{cantwell2015nektar++, karniadakis2013spectral}. 

\begin{figure*}
    \centering
    \includegraphics[width=0.7\textwidth]{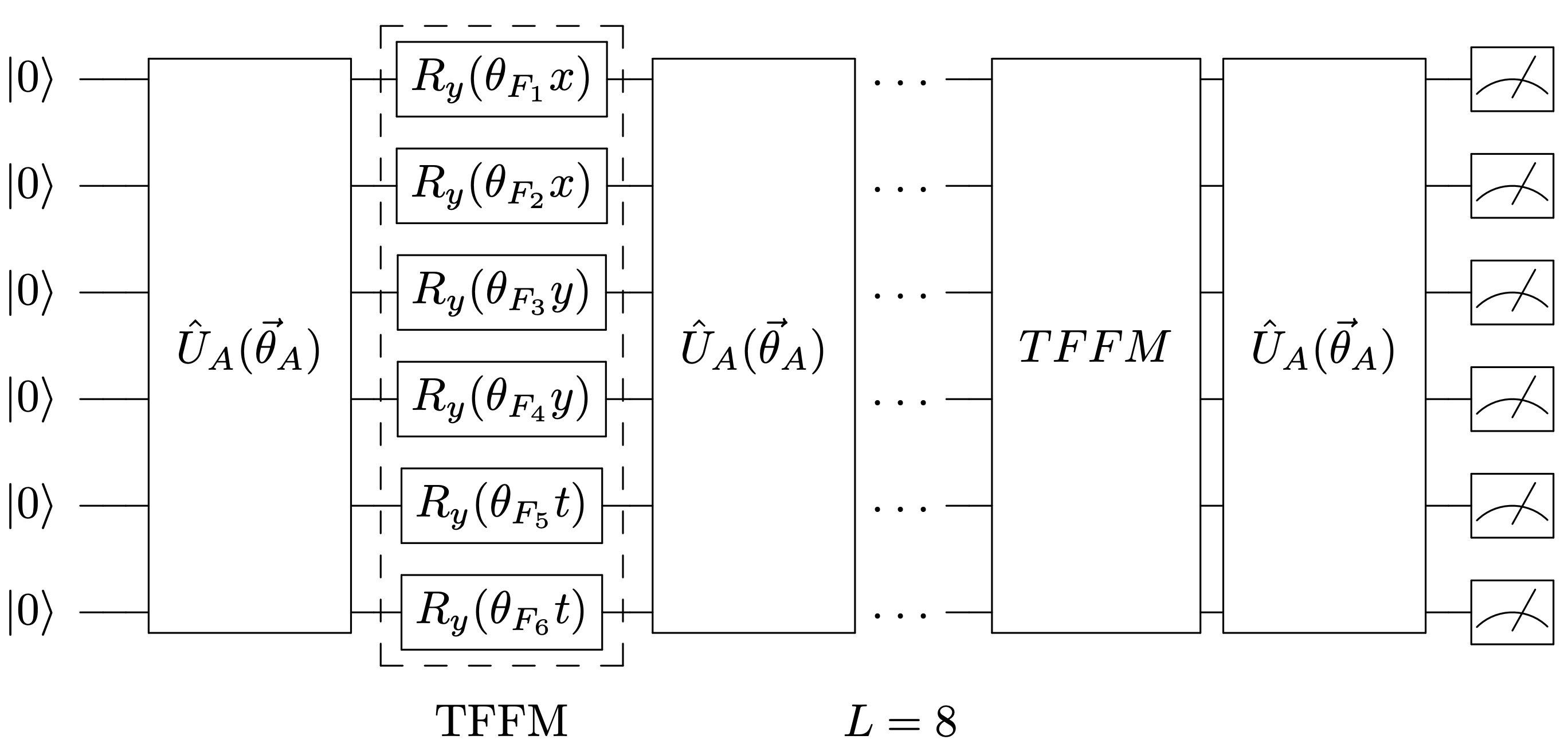}
    \caption{Circuit diagram of the trainable-frequency (TF) QNN architecture used in experiments solving the Navier-Stokes equations. The trainable-frequency feature map (TFFM, dashed box) contains the generator parameters $\theta_F$ that allow training of the underlying model frequencies. The TFFM is followed by $L=8$ ansatz layers, a data-reuploading and then a final ansatz layer before the qubits are measured. Here, we study both digital and digital-analog versions of such layered abstraction.}
    \label{fig:ns_circuit_diagram_new}
\end{figure*}

The quantum model is trained by minimising a loss $\mathcal{L} = \mathcal{L_{\text{PDE}}} + \mathcal{L_{\text{data}}}$. \revision{The equation loss $\mathcal{L_{\text{PDE}}}$ is given by}

\revision{
\begin{equation}
\mathcal{L_{\text{PDE}}} = \frac{1}{M}\sum_{i=1}^{M} \zeta_x(x_i, y_i, t_i)^2 + \zeta_y(x_i, y_i, t_i)^2, 
\label{eqn:pde_loss}
\end{equation}
}

\revision{where $x_i$, $y_i$ and $t_i$ are the coordinates of a collocation point $i$ in a total batch of $M$ collocation points. Importantly, here the terms $\zeta_x$ and $\zeta_y$ are evaluated with observables $\tilde{u}$ $\tilde{v}$ and $\tilde{p}$ predicted by the quantum model. Meanwhile, $\mathcal{L_{\text{data}}}$ is a supervised loss term}

\revision{
\begin{align}
\mathcal{L_{\text{data}}} = \frac{1}{M}\sum_{i=1}^{M} &[u(x_i, y_i, t_i) - \tilde{u}(x_i, y_i, t_i)]^2 \notag\\&+ [v(x_i, y_i, t_i) - \tilde{v}(x_i, y_i, t_i)]^2 \\&+ [p(x_i, y_i, t_i) - \tilde{p}(x_i, y_i, t_i)]^2\notag,
\label{eqn:data_loss}
\end{align}
}
\revision{
where the reference values $u$, $v$ and $p$ are given by the data set. Notably, the data set used in training contains only 1\% of the total points in the reference solution. This means that the remainder of the flow must be predicted by learning a solution that directly solves the NSEs.
}

Given the increased problem complexity compared to section \ref{sec:pop_results}, here we employ a more advanced quantum architecture. The overall quantum model consists of two QNNs\revision{. The output of the first QNN is the predicted pressure $\tilde{p}$. Meanwhile, the output of the second QNN is the predicted stream function $\tilde{\psi}$, a quantity from which the predicted velocities $\tilde{u}$ and $\tilde{v}$ can be obtained via the relations}

\revision{
\begin{align*}
\tilde{u} = \frac{\partial\tilde{\psi}}{\partial y} && \tilde{v} = -\frac{\partial\tilde{\psi}}{\partial x}.
\end{align*}
}
\revision{Computing the velocities this way} ensures that the mass continuity equation is automatically satisfied, which would otherwise require a third term in the loss function. \revision{Furthermore, the derivatives of $\tilde{p}$ and the derivatives of $\tilde{u}$ and $\tilde{v}$ required in Eq. (\ref{eqn:pde_loss}) can be computed from derivative quantum circuits of the first and second QNNs respectively as defined in the DQC algorithm~\cite{kyriienko2021solving}.}

The architecture of each QNN, consisting of $N=6$ qubits, is shown in Figure \ref{fig:ns_circuit_diagram_new}. First, a single ansatz layer $\hat{U}_A(\vec{\theta}_A)$ is applied, followed by a TFFM which encodes each dimension $(x, y, t)$ in parallel blocks. Each dimension of the TF encoding once again uses the simple parameterization $\hat{G}_\theta = \sum_{m=1}^N \theta_m\hat{Y}^m/2$, whilst the FF model uses the same generator without trainable parameters $\hat{G} = \sum_{m=1}^N \hat{Y}^m/2$. After the FM, a sequence of $L=8$ ansatz layers are then applied. Subsequently a data-reuploading feature map is applied, which is a copy of the TFFM block including sharing the parameters. Finally, the QNN architecture ends with a single ansatz layer. Overall, each QNN has a circuit depth of 52 and 180 trainable ansatz parameters.

\begin{figure*}
    \centering
    \includegraphics[width=\textwidth]{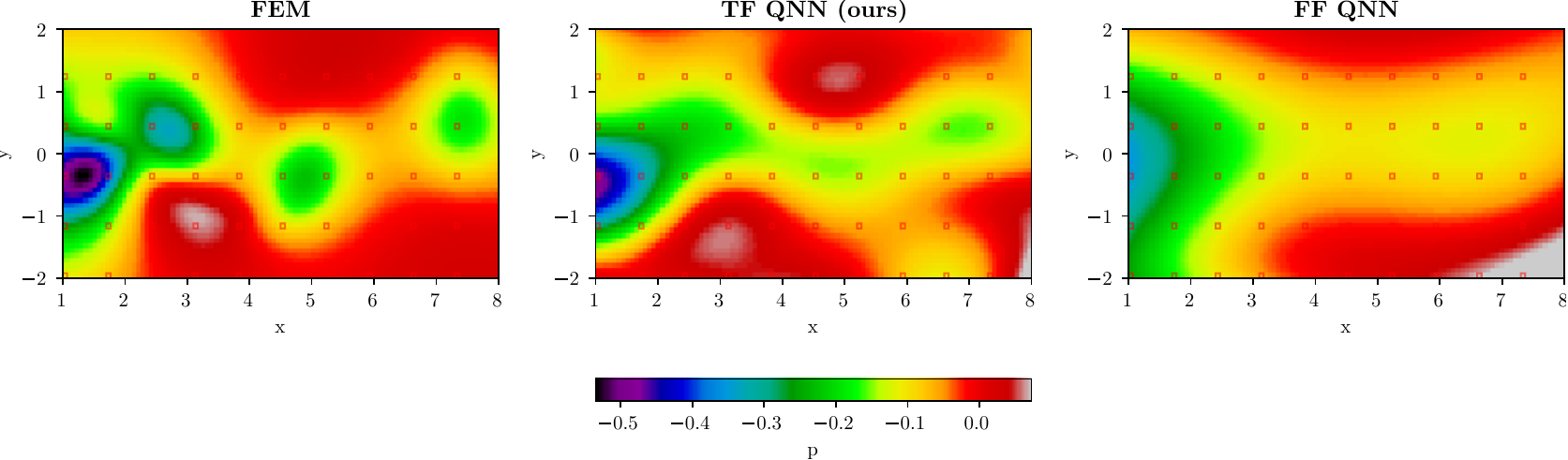}
    \caption{Pressure field at $t=3.5$ of the wake flow of fluid passing a circular cylinder at x=0. Left: reference solution obtained with the finite-element method (FEM) in~\cite{raissi2017physics}. Cells with red borders indicate the training data accessible to the quantum models, see main text. Middle: prediction of trainable-frequency (TF) QNN using generators $\hat{G}_\theta$. Right: prediction of fixed-frequency (FF) QNN using generators $\hat{G}$. The quantum circuits are based on a sliced Digital-Analog approach (sDAQC \cite{parrarodriguez2020digitalanalog}) suitable for platforms such as neutral atom quantum computers.}
    \label{fig:ns_p_visualisation_new}
\end{figure*}

Figure \ref{fig:ns_p_visualisation_new} gives a visualization of the results of this experiment, where the pressure field at a specific time is compared for different methods. Here the quantum models are trained for 5{,}000 iterations; more details can be found in Appendix \ref{app:hyperparameters}. The left panel shows the reference solution for the pressure $p$ at time $t=3.5$. Here, the $10\times5$ grid of cells with red borders correspond to the points that are given to the quantum models at each time step to construct $\mathcal{L_{\text{data}}}$, overall 1\% of the total $100\times50$ grid. The middle panel illustrates the prediction of the TF QNN. While the model does not achieve perfect agreement with the reference solution, it captures import qualitative features including the formation of a large negative-pressure bubble on the left and two additional separated bubbles on the right. By contrast, the FF QNN solution only correctly predicts the global background, unable to resolve the distinct different regions of pressure that form as fluid passes to the right. This demonstrates just how impactful TFFMs can be on the expressiveness of quantum models, even with limited width and depth. Further still, we show in Appendix \ref{app:ns_old_results} how deeper TF models can match the FEM solution whereas FF models cannot.

\begin{table*}[t]
\centering
\renewcommand{\arraystretch}{1.5}
\begin{tabular}{|c|c|c|c|c|c|c|}
\hline
       & \multicolumn{2}{c|}{u} & \multicolumn{2}{c|}{v} & \multicolumn{2}{c|}{p} \\
\hline
       & FF          & \textbf{TF}          & FF          & \textbf{TF}           & FF          & \textbf{TF}           \\
\hline
Min  & $16.8\pm0.3$       & \textbf{$\mathbf{12.8\pm1.9}$}          & $90.3\pm2.6$          & \textbf{$\mathbf{27.4\pm3.3}$}       & $62.9\pm1.9$         & \textbf{$\mathbf{51.0\pm6.0}$} \\
\hline
Max  & $18.3\pm0.4$      & \textbf{$\mathbf{14.2\pm3.0}$}          & $122.0\pm2.4$          & \textbf{$\mathbf{34.3\pm4.6}$}       & $78.3\pm1.8$        & \textbf{$\mathbf{62.8\pm10.3}$}          \\
\hline
Mean  & $17.5\pm0.3$      & \textbf{$\mathbf{13.2\pm2.0}$}         & $103.2\pm1.4$        & \textbf{$\mathbf{30.4\pm2.7}$} &       $73.2\pm1.9$           & \textbf{$\mathbf{57.0\pm3.7}$}          \\
\hline
\end{tabular}
\caption{MAERM error predicting $u$, $v$ and $p$ averaged over 10 runs. The minimum, maximum and mean are with respect to the 11 time points between $t=0$ to $t=5.5$.}
\label{table:ns_results_new}
\end{table*}

To quantify this benefit, the mean absolute error relative to the median (MAERM) $\frac{1}{\revision{\mathcal{N}}} \sum_{i=1}^{\revision{\mathcal{N}}} \left| \frac{{\hat{y}_i - y_i}}{{\tilde{y}}} \right|$ is calculated for each time step and observable, where the sum spans the $\revision{\mathcal{N}}$ spatial grid points. The results, presented in Table \ref{table:ns_results_new}, numerically demonstrate the improved performance of the TF model across all observables and time steps. Particularly notable is the large improved accuracy of the vertical velocity $v$.

Finally, it is worth considering how the inclusion of additional trainable parameters \revision{in the feature map of the TF model} affects the cost of training \revision{compared to FF models which have otherwise the same architecture}. \revision{When training using} a gradient-based optimizer such as Adam, \revision{regardless of the feature map chosen,} each training iteration requires the calculation of $\partial L/\partial\theta_i$ for all trainable parameters $\theta_i$ in the circuit. If one were to calculate these gradients on real quantum hardware, one would need to use the PSR. For the quantum architecture used in this section, all parameterized unitaries decompose into single-qubit Pauli rotations, such that only two circuit evaluations are required to compute the gradient of each parameter in $\vec{\theta}_A$ and $\vec{\theta}_F$. Thus, the factor $C_f = \frac{|\vec{\theta}_F| + |\vec{\theta}_A|}{|\vec{\theta}_A|}$ describes the additional circuit evaluations required to train the TF model, \revision{due to the additional parameters appearing in the corresponding feature map}. For the models used in this section, the cost factor is $C_f = \frac{12 + 180}{180} = 1.07$. This represents only a 7\% increase in the number of quantum circuits evaluated, a modest cost for the improvement in performance observed.

\section{Discussion}\label{sec:conclusion}

In this work we explore an extension of variational QML models to include trainable parameters in the data-encoding Hamiltonian, compatible with a wide range of models including those based on digital and digital-analog paradigms. As introduced in section \ref{sec:method}, when viewed through the lens of a Fourier representation, the effect of such parameters is fundamentally different from those in the variational ansatz, as they enable the frequencies of the quantum model to be trained. Furthermore, in section \ref{sec:pop_results} we showed how this leads to quantum models with specific spectral properties inaccessible to the conventional approach of tuning only the coefficients of fixed orthogonal basis functions. Finally, in section \ref{sec:ns_results} we demonstrated the benefit of TF models for practical learning problems, leading to a learned solution of the Navier-Stokes equations closer to the ground truth than FF models. 

We note that, in theory, a FF model could also achieve parity with TF models if it had independent control of the coefficients of each basis function. Given data with a minimum frequency gap $\Delta_{d, \text{min}}$ and spanning a range $r_d = |\Omega_{d, \text{max}} - \Omega_{d, \text{min}}|$, a quantum model with $\frac{r_d}{\Delta_{d, \text{min}}}$ fixed frequencies could span all modes of the data. In this case, such a model could even represent data with non-regularly spaced frequencies (e.g., $\Omega_d=\{1, 1.2, 3\}$ by setting the coefficients $\vec{c}_j = 0$ for $\vec{\omega}_j=\{1.1, 1.3, 1.4, ..., 2.9\}$). However, such a model would be exponentially costly to train, since independent control of the coefficients of the model frequencies would generally require $\mathcal{O}(2^N)$ ansatz parameters. It is for this reason that we present TF models as having a practical advantage, within the context of scalable approaches to quantum machine learning.

Looking forward, an interesting open question remains around the performance of other parameterizations of the generator. A notable instance of this would be $\hat{G} = \sum_{m=1}^N \text{NN}(\theta_F)\hat{Y}^m/2$, where the parameterization is set by a classical neural network. Interestingly, this has already been implemented in a different context in so-called hybrid quantum-classical networks~\cite{jaderberg2022quantum, mari2020transfer,liu2021hybrid,liu2022representation}, including studies of DQC~\cite{sedykh2023quantum}. The use of hybrid networks is typically motivated by the desire to relieve the computational burden from today's small-scale quantum models. Our work offers the insight that such architectures are actually using a classical neural network to set the frequencies of the quantum model. \revision{Promisingly, there is already early evidence to suggest that this scheme may lead to improved performance for classification~\cite{hur2023neural}.}

Compellingly, for many different parameterizations of TFFMs, a generator with regularly spaced eigenvalues is accessible within the parameter space. This is particularly true for the parameterizations studied in this work, which can be trivially realised as an orthogonal model when $\vec{\theta}_F=\mathds{1}$. This implies that, at worst, many classes of TF models can fall back to the behaviour of standard FF models. We find this, along with our results, a strong reason to explore in the future whether TF models could be an effective choice as a new default for variational quantum models.



\appendix

\section{Computing derivatives of trainable parameters in the generator}\label{app:tffm_derivatives}

Training a variational quantum circuit often involves performing gradient-based optimization against the trainable parameters of the ansatz. In this section we make clear how in the case of TF models, the trainable parameters in the generator can also be optimised in the same way. For gradient-based optimization, one needs to compute $\frac{\partial \mathcal{L}}{\partial \Vec{\theta}}$, for a suitably defined loss function $\mathcal{L}$ which captures the adherence of the solution $f (x)$ to the conditions set by the training problem.

Let us first define the state produced by a TF quantum model as 

\begin{equation}
  |f_{\vec{\theta}_F, \vec{\theta}_A}(x)\rangle = \hat{U}_A(\vec{\theta}_A) \hat{U}_F(x, \vec{\theta}_F) | 0 \rangle.
  \label{eq:quant_measured_f}
\end{equation}

The output of the quantum models used in the main text, given as the expectation value of a cost operator $\hat{C}$

\begin{equation}\label{eqn:1d_fourier_fm_with_thetas_output}
    f(x, \vec{\theta}_F, \vec{\theta}_A) = \langle f_{\vec{\theta}_F, \vec{\theta}_A}(x) | \hat{C} |f_{\vec{\theta}_F, \vec{\theta}_A}(x)\rangle,
\end{equation}

can then act as a surrogate for the target function $f$.

For a supervised learning (SVL) loss contribution we can define
\begin{equation}
\mathcal{L}_{SVL} = \frac{1}{M} 
\sum_i^M L( f(x_i, \vec{\theta}_F, \vec{\theta}_A), y_i),
\end{equation}
where $L$ is a suitable distance function. In a DQC setting, with each (partial, differential) equation embedded in a functional $F[\partial_X f(x), f(x), x]$ to be estimated on a set of $M$ collocation points $\{x_i\}$, one can define a physics-informed loss function as 

\begin{equation}
\mathcal{L}_{DQC} = \frac{1}{M} 
\sum_i^M L( F[\partial_X f(x_i), f(x_i), x_i], 0). 
    \label{eq:def_difflos}
\end{equation}

When optimising the loss against a certain variational parameter $\theta$, if $\theta \in \vec{\theta}_A $ is an ansatz parameter then $\frac{\partial \mathcal{L}}{\partial \theta}$ can be computed as standard with the PSR and generalised parameter-shift rules (GPSR) ~\cite{kyriienko2021solving, kyriienko2021generalized}.
If instead $\theta \in \vec{\theta}_F $, we can account for the inclusion of the feature $x$ using the linearity of differentiation

\begin{align}
    &  \frac{\partial \mathcal{L}_{SVL}}{\partial \theta} = \frac{1}{M}
    \sum_i^M \frac{d L}{d f} \frac{\partial f(x_i)}{ \partial \theta},
    \label{eq:svl_derivative_loss} \\
    & \frac{\partial \mathcal{L}_{DQC}}{\partial \theta} = \frac{1}{M}
    \sum_i^M \frac{\partial L}{\partial \theta} \left(F\left[ \partial_X f(x_i), f(x_i), x_i\right], 0 \right).
     \label{eq:dqc_derivative_loss}
\end{align}

Note that in Eq. \ref{eq:dqc_derivative_loss} we leave the right-hand side (RHS) as its implicit form, as it depends not only upon the actual choice of the distance $L$, as in Eq. \ref{eq:svl_derivative_loss}, but also upon the terms involved in the functional $F$.
In order to give further guidance on the explicit treatment of this latter case, we can split the discussion according to the various terms involved in the functional $F$, under the likely assumption that the problem variable(s) $X$ are independent from the variational parameter $\theta$:
\begin{itemize}
    \item Terms that depend solely on $x_i$ have null $\frac{\partial \cdot}{\partial \theta}$ and can be neglected.
    \item Terms containing the function $f$ itself can be addressed via the chain rule already elicited in Eq. \ref{eq:svl_derivative_loss}.
    \item Finally, terms depending on $\partial_X f(x_i)$ can be similarly decomposed as 
    \begin{equation}
    \frac{d^2 L}{df dX} \frac{\partial^2 f(x_i)}{\partial \theta \partial X} = \frac{d L}{d (\partial_X f)} \frac{\partial }{\partial X} \frac{\partial f (x_i)}{\partial \theta}
    \label{eq:mixed_terms}
    \end{equation}
    where the latter descends from the independence highlighted above, and the first term on the RHS can be simply attained from the (known) analytical form of $L$ and $F$.
\end{itemize}
    
Thus, following the chain rule, in all cases we obtain a dependency upon the term $\partial f(x_i) / \partial \theta$.

In terms of computing $\partial f(x_i) / \partial \theta$, we again omit the case where $\theta \in \Vec{\theta}_A$ as it is known from the literature \cite{mitarai2018quantum, kyriienko2021generalized}. For the specific case of $\theta \in \Vec{\theta}_F$ instead, let us first combine the unitary ansatz and the cost operator as $\hat{U}_A^{\dagger}(\vec{\theta}_A) \hat{C} \hat{U}_A(\vec{\theta}_A) \equiv \hat{C}_A(\vec{\theta}_A)$ for brevity. Rewriting Eq. \ref{eqn:1d_fourier_fm_with_thetas_output}, using the generic FM provided in Eq. \ref{eqn:fm} with a trainable generator $\hat{G}_m(\vec{\gamma}, \vec{\theta}_{F})$ and isolating the only term dependent on the $\theta_{\Tilde{m}}$ of interest, we get

\begin{align}
\label{eqn:fm_with_theta_expand}
    & \hat{U}_F(\vec{x}, \vec{\gamma}, \vec{\theta}_F) = e^{-i\hat{G}_1(\vec{\gamma}, {\theta}_1)\phi_1(\vec{x})} \otimes ... \nonumber \\ 
    & \otimes e^{-i\hat{G}_{\Tilde{m}}(\vec{\gamma}, {\theta}_{\Tilde{m}})\phi_{\Tilde{m}}(\vec{x})} \otimes ... \otimes e^{-i\hat{G}_M(\vec{\gamma}, {\theta}_M)\phi_M(\vec{x})} \\ 
    &\equiv \hat{U}_{FL} \otimes 
    e^{-i\hat{G}_{\Tilde{m}}(\vec{\gamma}, {\theta}_{\Tilde{m}})\phi_{\Tilde{m}}(\vec{x})} \otimes \hat{U}_{FR}.
\end{align}

Further simplifying the notation using $\hat{U}_{FR}| 0\rangle \equiv |f_{FR} \rangle$ and $\hat{U}_{FL}^{\dagger} \hat{C}_A \hat{U}_{FL} \equiv \hat{C}_{AF}$ produces
\begin{align}
\label{eq:contracted_elicited_f}
    &f(x, \theta_F, \theta_A) = \nonumber \\
    &\langle f_{FR} |
    e^{i\hat{G}_{\Tilde{m}}(\vec{\gamma}, {\theta}_{\Tilde{m}})\phi_{\Tilde{m}}(\vec{x})}
    \hat{C}_{AF}
    e^{-i\hat{G}_{\Tilde{m}}(\vec{\gamma}, {\theta}_{\Tilde{m}})\phi_{\Tilde{m}}(\vec{x})} 
    |  f_{FR} \rangle.
\end{align}

With the model expressed in terms of the dependency on the single FM parameter $\theta_{\Tilde{m}}$, we can now address computing $\partial f / \partial \theta_{\Tilde{m}}$:

\begin{align}
    &\frac{\partial f}{\partial \theta_{\Tilde{m}}} = \langle f_{FR} |
    e^{i\hat{G}_{\Tilde{m}}\phi_{\Tilde{m}}}
    \left[\phi_{\Tilde{m}}
    \frac{\partial \hat{G}_{\Tilde{m}}}{\partial \theta_{\Tilde{m}}}, \hat{C}_{AF}\right]
    e^{-i\hat{G}_{\Tilde{m}}\phi_{\Tilde{m}}} 
    |  f_{FR} \rangle,
    \label{eq:fm_theta_derivative}
\end{align}

where for simplicity we have omitted the parameter dependencies of $\phi_{\Tilde{m}}(\vec{x})$ and the $\hat{G}_{\Tilde{m}}(\vec{\gamma}, {\theta}_{\Tilde{m}})$ generator, and we have introduced the commutator notation $[\cdot, \cdot]$. 

Observing Eq. \ref{eq:fm_theta_derivative}, we are thus left with obtaining the partial derivative ${\partial \hat{G}_{\Tilde{m}}}/{\partial \theta_{\Tilde{m}}}$, and to this extent we distinguish three cases.  
(i) If $\hat{G}_{\Tilde{m}}(\vec{\gamma}, {\theta}_{\Tilde{m}}) = \gamma \theta_{\Tilde{m}} \hat{\sigma}^{\alpha}$, i.e. a single Pauli operator for a chosen $\alpha$ axis, then we can obtain the target derivative using the standard PSR.
(ii) When a similar dependency on the (non) trainable parameters holds, but we generalise beyond involutory and idempotent primitives, i.e.  $\hat{G}_{\Tilde{m}}(\vec{\gamma}, {\theta}_{\Tilde{m}}) = \gamma \theta_{\Tilde{m}} \hat{G}_{\Tilde{m}}$, we can instead rely on a single application of the GPSR to obtain ${\partial \hat{G}_{\Tilde{m}}}/{\partial \theta_{\Tilde{m}}}$. 
(iii) Finally, in the most generic case considered in this work, the spectral gaps of $\hat{G}_{\Tilde{m}}$ might depend non-trivially upon $\theta_{\Tilde{m}}$. In this last case, one should recompute such gaps for each new trained $\theta_{\Tilde{m}}$, in order to apply GPSR. 
Note, however, that one could always decompose $\hat{G}_{\Tilde{m}}(\vec{\gamma}, \theta_{\Tilde{m}}) = \sum_{i}^{I} \hat{P}_{i} (\Vec{\gamma}, \theta_{\Tilde{m}})$, i.e. a sum of Pauli strings $\hat{P}$. With this decomposition approach, ignoring any structure in $\hat{G}_{\Tilde{m}}$, at most $2I$ circuit evaluations would suffice to retrieve ${\partial \hat{G}_{\Tilde{m}}}/{\partial \theta_{\Tilde{m}}}$ \cite{hubregtsen2022single}.

\section{Quantum models and training hyperparameters}\label{app:hyperparameters}

\begin{figure*}
  \centering
  \begin{subfigure}[t]{0.5\textwidth}
    \vskip 0pt
    \includegraphics[width=\textwidth]{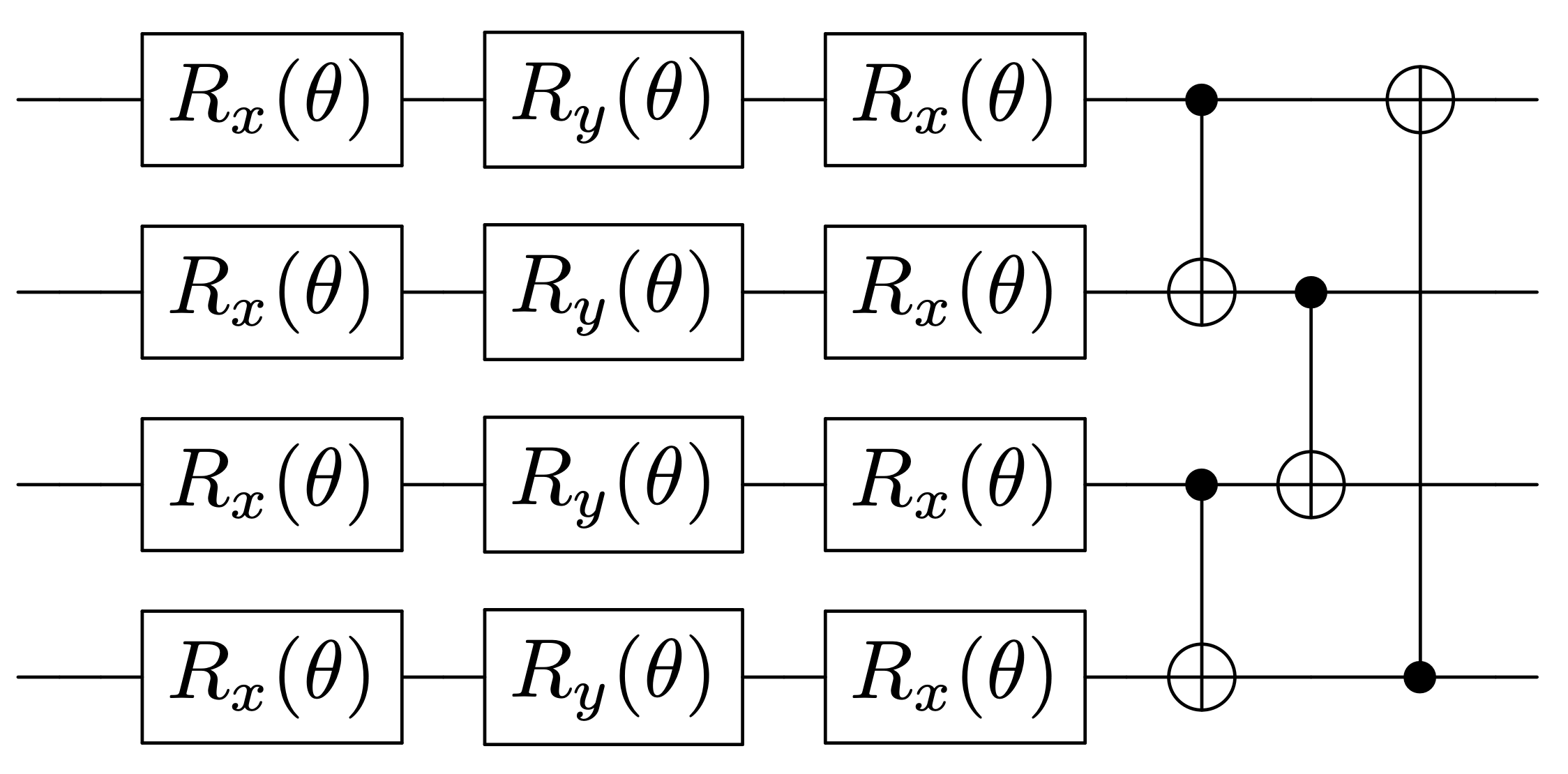}
    \caption{}
    \label{fig:hea_cx_n4}
  \end{subfigure}
  \hfill
  \begin{subfigure}[t]{0.45\textwidth}
    \vskip 0pt
    \includegraphics[width=\textwidth]{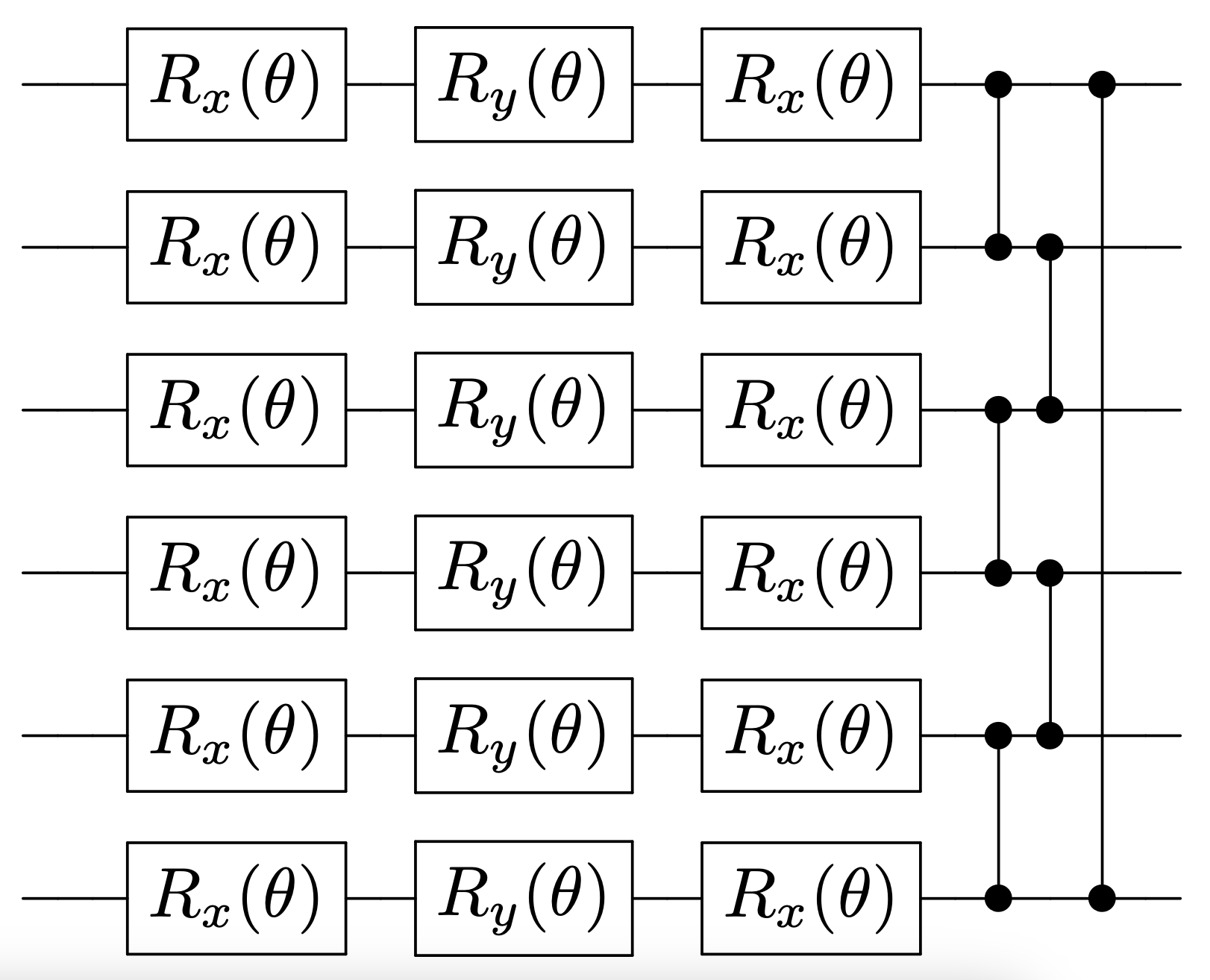}
    \caption{}
    \label{fig:hea_zz_n6}
  \end{subfigure}
  \caption{\revision{A single layer of of the different types of hardware-efficient ansatz (HEA) used in this work. All parameters $\theta$ are separate and independent.}}
  \label{fig:ansatz_diagrams}
\end{figure*}

For all quantum models used in this work, the variational ansatz is a variant of the hardware-efficient ansatz (HEA)~\cite{kandala2017hardware}, with entangling unitaries connecting the qubits in a ring topology \revision{as shown in Figure \ref{fig:ansatz_diagrams}. Here we study both digital and sDAQC \cite{parrarodriguez2020digitalanalog} versions of the HEA as shown in Figures \ref{fig:hea_cx_n4} and \ref{fig:hea_zz_n6} respectively. In the sDAQC approach, the entangling operations are fixed-duration Hamiltonian evolution of the form $\text{exp}(i \hat{n}_k\hat{n}_l\pi)$ between neighbouring pairs of qubits $(k,l)$.} After all FMs and ansatz layers are applied, a final state $\ket{\psi}$ is produced. The output (e.g., the predicted value of $y(x)$ or $\psi(x,y,t)$ or $p(x,y,t)$) of all models is obtained via the expectation value $\bra{\psi}\hat{C}\ket{\psi}$, where the cost operator $\hat{C} = \sum^N_{m=1} \hat{Z^m}$ is an equally-weighted total magnetization across the $N$ qubits. This combination of constant depth ansatz and 1-local observables is known to avoid cost function induced barren plateaus~\cite{cerezo2021cost}.
\\\\
All models in this work were trained using the Adam optimizer.
\\\\
In Figure \ref{fig:spectral_spacing_a} the model was trained with hyperparameters: qubits $N=4$, \revision{$L=4$ layers of the ansatz in Figure \ref{fig:hea_cx_n4}}, training iterations $N_i = 2{,}000$, batch size $b_s=1$, learning rate $\text{lr}=10^{-3}$. 
\\\\
In Figure \ref{fig:spectral_spacing_b}, both models were trained with: $N=4$, \revision{$L=4$ layers of the ansatz in Figure \ref{fig:hea_cx_n4}}, $N_i = 6{,}000$, $b_s=2$, $\text{lr}=10^{-3}$. 
\\\\
In Figure \ref{fig:spectral_richness_results} all models were trained with: $N=4$, \revision{$L=8$ layers of the ansatz in Figure \ref{fig:hea_cx_n4}}, $N_i = 4{,}000$, $b_s=2$, $\text{lr}=10^{-3}$. 
\\\\
In Figure \ref{fig:ns_p_visualisation_new} and Table \ref{table:ns_results_new}, all models were trained per QNN with: $N=6$, \revision{$L=10$ layers of the sDAQC ansatz in Figure \ref{fig:hea_zz_n6}}, $N_i = 5{,}000$, $b_s=600$, $\text{lr}=10^{-2}$.
\\\\
In Appendix \ref{app:ns_old_results}, Figure \ref{fig:ns_p_visualisation} and Table \ref{table:ns_results}, all models were trained per QNN with: $N=4$, \revision{$L=64$ layers of the ansatz in Figure \ref{fig:hea_cx_n4}}, $N_i = 5{,}000$, $b_s=600$, $\text{lr}=10^{-2}$.

\section{Navier-Stokes results in the over-parameterized regime}\label{app:ns_old_results}

\begin{figure*}
    \centering
    \includegraphics[width=\textwidth]{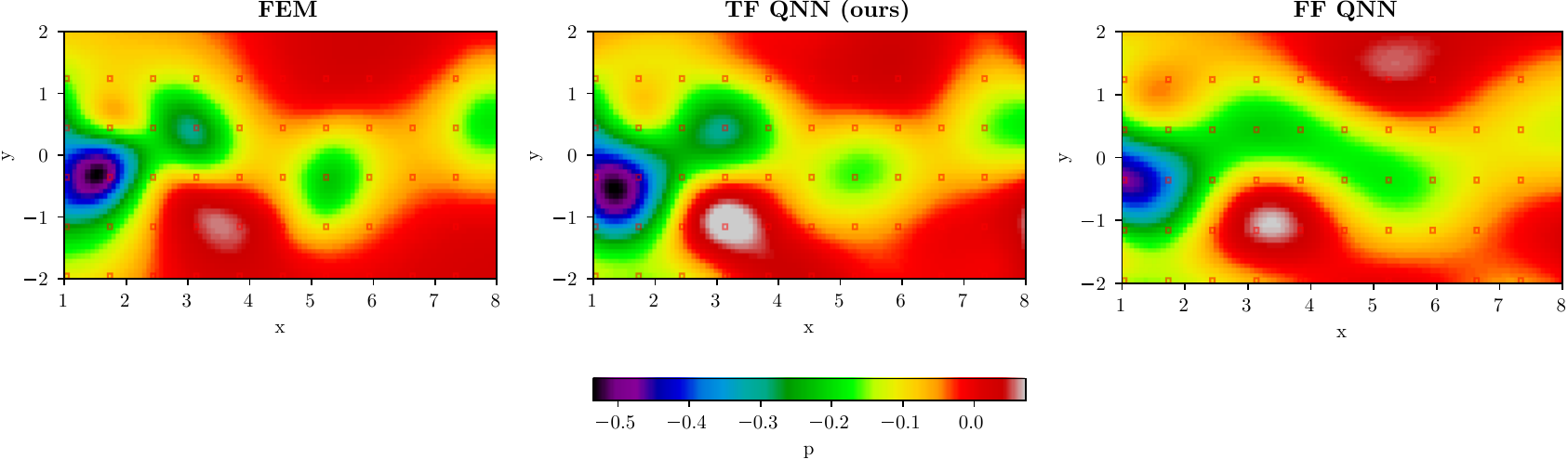}
    \caption{Pressure field at $t=4.0$ of the wake flow of fluid passing a circular cylinder at x=0. Left: reference solution obtained by FEM. Middle: prediction of the over-parameterized TF QNN model with $L=64$. Right: prediction of equivalent FF QNN.}
    \label{fig:ns_p_visualisation}
\end{figure*}

In this section we repeat the experiments in section \ref{sec:ns_results} using an over-parameterized model, such that the number of trainable parameters is larger than the dimension of the Hilbert space~\cite{you2022convergence}.  Furthermore, here we use a serial TFFM in which each dimension $(x, y, t)$ is encoded serially in separate blocks, separated by an ansatz layer which acts to change the encoding basis to avoid loss of information. This is theoretically preferential to the parallel encoding strategy since it produces a quantum model with more unique frequencies. The TFFM uses the simple parameterization $\hat{G}_\theta = \sum_{m=1}^N \theta_m\hat{Y}^m/2$ for each dimension, whilst the FF model uses the same generator without trainable parameters $\hat{G} = \sum_{m=1}^N \hat{Y}^m/2$. After the FM, the model has $L=64$ ansatz layers bisected by a data-reuploading FM. In total, each QNN has 804 trainable ansatz parameters.

Figure \ref{fig:ns_p_visualisation} presents a visualization of the experiment's results, evaluating the pressure field at a specific time. In this deeper regime, the TF QNN achieves excellent agreement with the reference solution, successfully capturing more features such as the presence of two interconnected negative-pressure bubbles on the left. In contrast, despite its increased depth, the FF QNN solution is only approximately accurate and fails to correctly identify the separation between the pressure bubbles in the middle and right. The numerical performance of the models is given in Table \ref{table:ns_results}.

\begin{table*}[t]
\centering
\renewcommand{\arraystretch}{1.5}
\begin{tabular}{|c|c|c|c|c|c|c|}
\hline
       & \multicolumn{2}{c|}{u} & \multicolumn{2}{c|}{v} & \multicolumn{2}{c|}{p} \\
\hline
       & FF          & \textbf{TF}          & FF          & \textbf{TF}           & FF          & \textbf{TF}           \\
\hline
Min  & $7.2\pm0.5$       & \textbf{$\mathbf{5.3\pm0.6}$}          & $23.1\pm0.8$          & \textbf{$\mathbf{16.8\pm0.9}$}       & $39.2\pm1.9$         & \textbf{$\mathbf{31.2\pm3.9}$} \\
\hline
Max  & $8.5\pm0.6$      & \textbf{$\mathbf{8.1\pm1.3}$}          & $30.0\pm0.7$          & \textbf{$\mathbf{22.3\pm1.8}$}       & $54.2\pm4.1$        & \textbf{$\mathbf{47.4\pm5.7}$}          \\
\hline
Mean  & $7.7\pm0.3$      & \textbf{$\mathbf{6.2\pm0.4}$}         & $26.0\pm0.5$        & \textbf{$\mathbf{18.9\pm0.7}$} &       $45.2\pm3.3$           & \textbf{$\mathbf{36.8\pm2.2}$}          \\
\hline
\end{tabular}
\caption{MAERM error predicting $u$, $v$ and $p$ averaged over 10 runs. The minimum, maximum and mean are with respect to the 11 time points between $t=0$ to $t=5.5$.}
\label{table:ns_results}
\end{table*}

\bibliography{bibliography.bib}

\end{document}